\documentclass[11pt,reqno]{amsart}
\usepackage{geometry}

\usepackage{amsmath}
\usepackage{amsfonts}
\usepackage[all]{xy}
\usepackage{amssymb}
\usepackage{xcolor}
\usepackage{dsfont}
\usepackage{color}
\usepackage{cite}
\usepackage{graphicx}
\usepackage{braket}
\usepackage{slashed}
\usepackage{float}
\usepackage{braket}
\usepackage{color}
\usepackage{upgreek}
\usepackage{hyperref}
\usepackage{physics}
\usepackage{tikz}
\usepackage{tikz-cd}

\hypersetup{
    colorlinks,
    linkcolor={blue!50!red},
    citecolor={orange!70!red},
    urlcolor={blue!80!black}
} 

\def\ab{{\alpha\beta}}

\def\a{\alpha}
\def\b{\beta}
\def\g{\gamma}

\def\R{\mathbb{R}}
\def\constant{\eta_{\a \b \g}}

\def\be{\begin{equation}}
\def\ee{\end{equation}}

\def\R{\mathds{R}/\{0\}}
\def\Rd{\mathds{R}}
\def\Co{{\mathds{C}P^1}}
\def\Con{{\mathds{C}P^n}}
\def\Hn{{\mathds{H}P^n}}
\def\Z{\mathds{Z}}

\def\U{\mathcal{U}}
\def\Ua{\mathcal{U}_\alpha}
\def\Ub{\mathcal{U}_\beta}
\def\Ug{\mathcal{U}_\gamma}
\def\world{\varGamma}
\def\cech{\v{C}ech }
\def\F{\mathfrak{F}}
\def\chain{C^p(\U, \F)}
\def\semicolon{{; }}
\def\double{{\mathbf {C^p(\U,\F,\Omega^q)}}}

\newtheorem{theorem}{theorem}

\begin{document}

\title{Local aspects of topological quantization and the Wu-Yang Monopoles}
\author{Aayush Verma}
\maketitle
\begin{abstract}
    In this paper, we review how local potentials arise in the Wu-Yang topological quantization. We also discuss the isomorphism between the de Rham cohomology classes and \v{C}ech cohomology classes in such topological quantization. We also emphasize the importance and application of local and global information in gauge theories.
\end{abstract}

\vspace{3in}
\begin{flushleft}
   {\sf \today}
\end{flushleft}

\newpage

{\hypersetup{linkcolor=black}
              \tableofcontents}
\section{Introduction}

The study of monopoles first appeared in \cite{dirac1931quantised}, in which Dirac proposed a quantization condition that implies the quantization of electric charge $e$ in the presence of magnetic monopoles of strength $G$. Dirac monopoles are defined for symmetric Maxwell fields. In particular, we put $G$ at the origin which produces a magnetic field
$
\mathbf{B} = \frac{G}{\rho^2}\hat{e}_\rho
$
on $\mathds{R}/\{0\}$. It is important to note that the fundamental group $\pi_1(\R)$ is trivial which means that $\R$ is a simply connected topology. An equivalent statement is that two paths are homotopy invariant and can be contracted to a point in $\R$. This property is necessary to realize that there exists smooth 1-form potential ${A}$ with $dA = B$. However, such potentials are hard to define on $\R$ for monopoles, as we will see later.

In order to have a viable topology, Dirac suggested a string $D_s$ which originates from the origin and moves to infinity without intersecting itself. For such strings, we can define its complement as an open subset $U$ in $\mathds{R}^3$ which has $\pi_2(U) =0$. This suggests that no sphere in $U$ contains the points of $D_s$. Moreover, $U$ is simply connected as well. So we can take a loop around $D_s$, continuously lift around the origin, and shrink it to the point. We can imagine two Dirac strings $D_{s+}$ and $D_{s-}$ with a common origin and we get two open sets $U_+$ and $U_-$ which can cover the $\R$. We can now define two potentials (up to scalar multiple) on each open set which we denote as $A_+$ and $A_-$. As expected we find that $A_+$ and $A_-$ do not agree on $U_+ \cap U_-$ which is just $\Rd/D_{s{\pm}}$. (If they had agreed, that means there exists an $A$ globally on $\R$ without difficulties.) For this space configuration, $F$ is exact since
\begin{equation}
H^2_{dR}(R/D_{s\pm})=0
\end{equation}
and thus it resolves, somewhat, the problem of singularity. However, this is not the nicest solution.

The `Dirac quantization condition' is a result of the quantum mechanical nature of the phase factor. Precisely, it is given for an integer $n$
\begin{equation}\label{quantization}
qG = \frac{1}{2}n
\end{equation}
which is a direct manifestation of a Hopf bundle, which we will discuss later. In \eqref{quantization}, $G$ is the monopole strength.\footnote{We have chosen $G$ as a symbol instead of $g$ here to not get confused with gauge transformations used in this paper by notation $g_{\a \b}$.} It is also interesting to note that a generalized Chern-Gauss-Bonnet theorem also implies this quantization condition.

Now we argue why there is not a ``well-defined'' non-singular potential $A$ for $B$. For this, we take a 2-sphere $S^2$ on $\R$ and divide the 2-sphere $S^2$ into two manifolds given by $R_+$ and $R_-$. We assume that there exists a non-singular potential $A$ on $S^2$ with $dA = B$. Provided $B$, we simply have
\begin{align}
\begin{split}
\iint_{S^2}B \cdot dS^2 &= \iint_{S^2} \left( \frac{g}{\rho^2 }\hat{e}_\rho \right) \cdot \hat{e}_\rho dS^2 \\
&= \frac{g}{R^2} \iint_{S^2} dS^2 \\ 
&= 4\pi g
\end{split}
\end{align}
which seems to be the right answer. However, we now do the same integral using Stoke's theorem. We should be able to do it assuming that $A$ is a smooth 1-form. We see that
\begin{align}
\begin{split}
\iint_{S^2}B \cdot dS^2 &= \oint_C A \cdot dr+ \oint_{-C} A \cdot dr \\
&=0
\end{split}
\end{align}
which is a contradiction. To rescue from this precise singularity\footnote{This singularity is not really physical, since one can define a global form $F$ without any singularity. This is also one of the reasons why Dirac's string formalism and Wu-Yang formalism are equivalent. See \cite{brandt1977dirac} for more.} of $A$ on $B$, Wu and Yang came up with a gauge interpretation (or `loops' that we will see) \cite{wu1975concept,brylinski2007loop}. This is one of the central topics in this paper. We must realize that what we are doing is essentially a localization. For instance, if $\bigcup_i \alpha_i = X$ for a space $X$ and we can describe a continuous function $f \colon X \rightarrow \Rd$ where we can check the agreement for two by two intersections. The localization of potential is the only way to define non-singular potentials, however, globally there exists still a singular potential.

Before discussing the Wu-Yang gauge approach, we must look into the Hopf bundle. It is interesting to observe that the connection on the Hopf bundle ($\Omega \colon S^3 \rightarrow S^2, \Omega$ being a functor in this paper) describes monopoles that we have just described \cite{ryder1980dirac,trautman1977solutions}.

\begin{figure}[ht!]
    \centering
    \includegraphics[scale=.6]{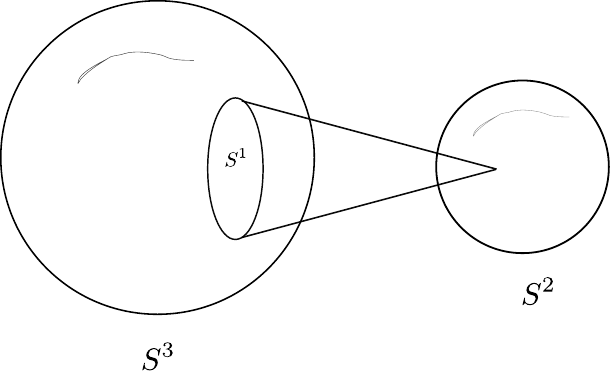}
    \caption{Hopf Fibration}
\end{figure}

The study of homotopy groups of spheres is a useful task. One such is homotopy groups of 2-sphere $\pi_i(S^2)$. To understand the Hopf bundle, we define a group action $U(1)$ on 3-sphere $S^3$ defined using complex coordinates $z_0,z_1$. The group action for $u \in U(1), ||u||=1$ is given by
\begin{equation}
u (z_0,z_1) = (uz_0,uz_1)
\end{equation}
and the quotient\footnote{A natural question to ask in what kinds of Maxwell's equations we get using $\Con$ or perhaps $\Hn$. For ${\mathds{C}P^2}$ we find that we get an electromagnetic instanton \cite{trautman1977solutions}. Similarly, we can move higher in dimensions for either finding higher dimensional solutions to Maxwell fields or quaternions fields. Hopf bundle description comes handy in here.} of $S^3$ by the action is $\Co$. The fibers of $U(1)$ in $S^3$ are $S^1$. Now the $\Co$ is equivalent to Riemann sphere $S^2$, hence $\Co \simeq S^2$. The Hopf bundle now becomes $S^1 \rightarrow S^3 \xrightarrow {\Omega} S^2$. One can also prove that $\pi_i (S^3) = \pi_i (S^2)$. We can now define the curvature of the connection on $S^3$ which is $S^1$ bundles over $S^2$ as 
\begin{equation}
F = \frac{1}{2} \sin \theta\ d \phi \wedge d \theta
\end{equation}
which is, when extended to Minkowski spacetime, just the field strength for our monopoles of strength $g = \frac{n}{2q}$. This is one of the many definitions of the Hopf bundle, we have described it in a way that is useful to us, namely in terms of homotopy groups. And indeed, $\R$ is homotopy equivalent to $S^2$.

We can also argue using this terminology why there is no singular potential on $\R$. This is because of the fact that Hopf bundle is non-trivial 
\be 
\pi_1(S^2 \times S^1) \approx \Z \neq \pi_1(S^3)
\ee
where $\pi_1(S^3)$ is trivial and $\pi_1(S^2)=\Z$. It is worth noting that when we apply the Wu-Yang method, to be described in the next subsection, the charge quantization Eq.~\eqref{quantization} is provided by $\pi_1(U(1))= \mathds{Z}$.

\subsection{The Problem with the Dirac String} {\it This is not related to the rest of the paper.}
Dirac had suggested that due to the singularity in the potential forms, we must use a Dirac string as we mentioned previously. It was suggested that the Dirac string is not physical and a mathematical workaround. However, it is not the best way to handle the singularity. We do not need Dirac strings in `t hooft-Polyakov monopoles as there are no singularities in those non-abelian monopoles. For a good exposition of Dirac string see \cite{Heras:2018uub}. The essential problem with Dirac string, which is that it does not completely eliminate the problem of finding a global 1-form potential, is well-known. We will find that some theorems of algebraic topology obscure us from finding a global potential over the manifold. Recently, it was suggested by the authors of \cite{Gonuguntla:2023cik} that there is a hidden field momentum contribution from Dirac string which violates the center of energy theorem \cite{coleman1968origin}. 

The author's point is as follows. We start with a simple monopole placed at the origin of ${\mathbb R}^3$ such that the magnetic charge and electric charge are at rest. The field momentum of the electric field by this monopole has two components which are Coulomb's term and Dirac's string term. There is a non-zero mechanical field momentum contribution from the interaction of magnetic charge and electric field due to the inclusion of Dirac's string which is not vanishing at all. See \cite{Gonuguntla:2023cik} for the discussion on this term. It was suggested in same that there are two takeaways from this non-trivial mechanical field momentum 1) the first is to say that the center of energy theorem is wrong which implies that this term is an error and 2) the second is to believe in the center of energy theorem and accept this term as a real contribution which implies that Dirac's string is real and must be physical even though how infinitesimally thin we believe it to be. However, then it becomes a system in which the electric charges generate a monopole-like magnetic field with a solenoidal magnetic flux \cite{Gonuguntla:2023cik}. Also, see this paper \cite{Gockeler:2023ugk} which is a comment on this violation. However, none of this will affect any of our discussions on what to follow.

\subsection{Greub-Petry-Wu-Yang Quantization}\label{1.1}
In this subsection, we recall the Wu-Yang method \cite{wu1975concept} (see also Greub-Petry \cite{greub1975minimal}) for describing monopoles with a charted 2-sphere. That is equivalent to the connection defined on Hopf bundle $S^3 \rightarrow S^2$. While doing so, we want to achieve potentials that are not singular and can be described using gauge theory\footnote{For a principle G-bundle, the global gauge group is defined as the bundle of automorphism. Local gauge group is the group of gauge transformation which trivially means changing the variables of the theory. Throughout the paper, we will be mainly talking about the local gauge group of gauge potentials. For that, one requires a trivialization of the manifold.}. We can achieve it but at the cost of the non-global theory of potential, for the reasons we have described above. We will set up the structure in a manner that would be fruitful in the context of the paper.

Originally, the Wu-Yang structure was constructed to provide meaning to the non-integrable phase factor. The idea is to take a 2-sphere and cover it with open covers $\mathfrak{U} = \{\mathcal{U}_i \} $. These open covers would be different from an open ball and we will call them `good' open covers. The last point enables us to use the Poincare lemma in every non-empty overlap, see \cite{bott1982differential}. For simplicity, we chart the sphere with two open covers $\U_\alpha$ and $\U_\beta$. However, it does not matter which open covers we need to use and one can use $n$ covers at a time. The properties of topological invariants do not depend on the choice of covers.\footnote{Such independence arises because of the nature of covers which are diffeomorphic to open ball. The intersection region, like $\Ua \cap \Ub$ is also diffeomorphic to an open ball, and any finite intersection is contractible. There also does not exist a unique point in the intersection of a worldline of a particle going through the overlap.} On each patch, we can associate a vector potential $A_{\alpha}, \alpha \in \Lambda$, which are 1-forms in de Rham complex. We associate $A_\alpha$ and $A_\beta$ to $\U_\alpha$ and $\U_\beta$ respectively. As one can check $A_{\alpha}$ are singularity free and gives $F = dA_\alpha$ for some region with boundary $V_i$. $F$ is also the curvature of $U(1)$ connection and it is globally invariant unlike potential forms.

Under simple circumstances of \cite{wu1975concept}, for two patches these potentials are
\begin{equation}
A_\alpha = \frac{G}{r\sin \theta}(1-\cos \theta)
\end{equation}
\begin{equation}
A_\beta = \frac{-G}{r \sin \theta}(1+\cos \theta)
\end{equation}
where $0 \leq \theta \leq \pi$ and $r >0$. In an non-empty overlap the region $\U_\alpha \cap \U_\beta$, $A_\alpha$ and $A_\beta$ are related by a gauge transformation
\begin{equation}\label{gauge}
A_\alpha \rightarrow g^{-1} A_\beta\ g
\end{equation}
and because this gauge transformation must be single-valued, due to Eq. \eqref{quantization}, there does not exist an overlap region that provides meaning to phase factors $\psi_i$ if Eq. \eqref{quantization} is unsatisfied. The field strength is invariant under a gauge transformation, up to gauge distortions. One could describe a similar gauge transformation 
\begin{equation}
A_\alpha \rightarrow \bar{g}^{-1}A_\beta\ \bar{g}
\end{equation}
but there exists a non-singular map $\lambda \colon g \rightarrow \bar{g}$ and quantization is invariant. Later, it will be evident to us that $g$ depends on the trivialization of the manifold.
In this way, we can describe a local theory of Dirac monopoles, without introducing strings. These are singularities free as one can see using Stoke's theorem. We stress again that the solution is invariant for any number of open covers, up to a topological constant.

In the language of \textbf{de Rham cohomology}, which is going to be the standard language henceforth, $A_{\alpha}$ is a $1$-form defined on $\U_i$. Our de Rham operator $d$ is defined nilpotent $d^2=0$ and forms a cochain complex of objects $p$-forms $\Omega^p_M$. de Rham cohomology $H^p_{dR, M}$ is defined as a set of closed forms modulo exact forms in $\Omega^p_M$ for defined on some manifold $M$. Since $F = dA$, it is a closed $2$-form in $\Omega^2_M$. What is the meaning of gauge transformation and quantization in cohomology theory?

The answer to the quantization interpretation is straightforward \cite{alvarez1985topological} and there is a thoughtful way to do it. We again take a 2-sphere and chart with three covers $\Ua, \Ub$ and $\Ug$. Similarly, we define 1-form gauge potentials $A_\alpha$, $A_\beta$ and $A_\gamma$ on each patch with gauge transformation between them in each non-empty overlap as defined in \eqref{gauge}. Now we can take a particle and describe its worldline in $M$ as $\varGamma$. As it goes through $S^2$, it acquires a correction to its Lagrangian of form
\begin{equation}\label{lagrangian}
\int_{\world} A.
\end{equation}
It should be noted that these are not the only corrections, for instance, particles can also interact with other gauge fields. We include in \eqref{lagrangian} only those interactions which are of `topological interest' to us. Since there does not exist a global vector potential $A$, we must refer to the Wu-Yang structure. The trajectory of $\world$ goes through each patch in $S^2$. Which point we choose in any finite overlap of $\Ua, \Ub, \Ug$ to pass $\world$  does not matter because we have used good finite covers \cite{bott1982differential,alvarez1985topological}. By a slight abuse of notation, we will describe region\footnote{The boundary condition can be described as $\partial \U(\a \b) = \U_\a - \U_\b$.} $\U(\alpha \beta) = \Ua \cap \Ub$ and likewise we define $\U(\beta \gamma)$ and $\U(\gamma \alpha)$. We can now define a point $P_{\U(\alpha \beta)}$ in $\U(\alpha \beta)$. Similarly, points $P_{\U(\beta \gamma)}$ and $P_{\U(\gamma \alpha)}$. It should be noted that these points can be picked up arbitrarily in each overlap and its position does not determine the final solution.
\begin{figure}[t]
    \centering
    \label{cover}
    \includegraphics[scale=.9]{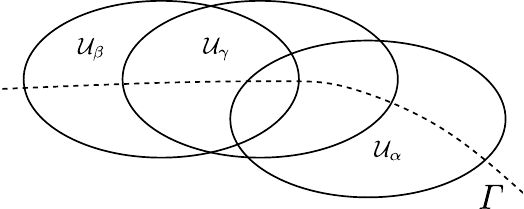}
    \caption{The worldline $\world$ is the dotted line passing through the overlaps between $\Ua, \Ub,$ and $\Ug$.}
\end{figure}

The path of $\world$ goes through our defined points in $P_{\U(\alpha \beta)}, P_{\U(\beta \gamma)}, P_{\U(\gamma \alpha)}$ and transition between potential $A_{\alpha}$ is described by the gauge transformation (or transition function) $g$. For convenience, we can define a map
\begin{equation}
g_{\a \b} \colon A_\a \rightarrow A_\beta
\end{equation}
and the co-boundary is
\begin{equation}\label{diffgauge}
dg_{\a \b} = A_{\a} - A_{\b}.
\end{equation}
If $g_{\a \b}$ is $n$-form, then $dg_{\a \b}$ is $(n+1)$-form. Since $A_\a - A_\b$ is a 1-form, $g_{\a \b}$ is a zero-form object. Transition functions $g_{\a \b}$ are anti-symmetric so $g_{\a \b} = - g_{\b \a}$. For 3-folds of manifold, see fig.~\eqref{cover}, the total contribution $\eqref{lagrangian}$ is given by the vector potentials and their gauge transformation. We can write the contribution simply \cite{alvarez1985topological}
\begin{align} \label{longlang}
I &= \int_{\Ua} A_\a + \int_{\Ub} A_\b + \int_{\Ug} A_{\gamma}
 + g_{\a \b} (P_{\U(\alpha \beta)}) + g_{\b \g} (P_{\U(\b \g)}) \\ &+ \{ g_{\a \b} (P_{\U(\g \a)}) + g_{\b \g} (P_{\U(\g \a)}) + g_{\g \a} (P_{\U(\g \a)})\}
\end{align}
which has numerous line integrals and gauge transformations. But there is a ``constant'' object that appeared at the end of the action \eqref{longlang} which is the piece in the curly bracket. To understand this constant, let us visit \eqref{diffgauge} and write further 
\begin{equation}
dg_{\b \g} = A_{\b} - A_{\g}
\end{equation}
\begin{equation}
dg_{\g \a} = A_{\g} - A_{\a}
\end{equation}
and all these equations give
\begin{equation}
d(g_{\a \b}+g_{\b \g}+g_{\g \a}) = 0
\end{equation}
which gives
\begin{equation}\label{constant}
g_{\a \b}+g_{\b \g}+g_{\g \a} \in \mathbb{R}
\end{equation}
that means, the piece is indeed a constant. It must be noted that it is constant also because it is a closed zero-form object, more on this later. Because of the Poincare lemma, we can give a more precise meaning of \eqref{constant}. In particular, one understands that 
\begin{equation}\label{consistency}
g_{\a \b}+g_{\b \g}+g_{\g \a} = \eta_{\a \b \g}
\end{equation}
is a constant only over the entire triple overlaps $\Ua \cap \Ub \cap \Ug$. It is interesting to note that since \eqref{constant} is a constant, it is irrelevant to define the coordinates in the last piece in \eqref{longlang}. We will interpret eq.~\eqref{constant} as a non-trivial cocycle condition in sec.~\ref{second}.

What if we had used two covers instead of three covers? It does not matter. As one can see, if one does two covers situations, there is no constant in action. In three covers cases, there is a constant. We can conclude that solutions of any number of covers are equivalent, up to a topological ambiguous constant. When we are solving classical equations, this constant can be ignored, however, in quantum mechanics, this constant leads to inconsistencies unless defined exactly.

In computing the line integral in eq.~\eqref{longlang}, the phase factor is found to be ambiguous for a Euclidean propagator in the form of $exp(i \eta_{\a \b \g})$. To be consistent, we must require every phase factor to be one, which means
\begin{equation}
 \constant = 2\pi \epsilon_{\a \b \g}
 \end{equation} 
 where $\epsilon_{\a \b \g} \in \mathbb{Z}$. A precise meaning of this constant exists in de Rham-Cech cohomology and also in the context of \cech cohomology in some presheaf that we will see later in sec.~\ref{second}.

Let us now turn to compute magnetic flux using Stoke's law. That requires us to subdivide a manifold into region $V_{\a \b \g} = \Ua \cap \Ub \cap \Ug \cap S^2$ and in the triple overlaps, the total flux determined by eq.~\eqref{constant}. Indeed, one can check that
\begin{align}\label{flux}
\int_{S^2} F &= \sum_{V_{\a \b\ \g}} \constant\\
& = 2\pi \sum_{V_{\a \b \g}} \epsilon_{\a \b \g}
\end{align}
and thus magnetic flux is determined by this constant. This is the famous Dirac's quantization condition. Since the singularity in $A$ is not physical, Dirac's string quantization and Wu-Yang quantization are equivalent \cite{brandt1977dirac}. So as long as we stick to Greub-Petry-Wu-Yang quantization anything that we describe in this paper should be equivalent to a physical Dirac monopole.

The local aspects of gauge potentials were used in the overall process of quantization. An important requirement in this topological quantization is the use of good covers $\U$. We will provide more meaning to all the objects that appeared in this section in the later part of this paper. 

\section{Monopoles and de Rham-\cech Cohomology}\label{second}
What we have observed so far is that the use of finite `good' covers can be helpful in understanding topological quantization. We already have described a cochain complex operated by a differential operator $d$, which is the de Rham operator. This operator induces cohomology in this cochain, namely\footnote{A notation clarification\semicolon we conventionally always use subscript for homology group and superscript for cohomology group. In this paper, we are only concerned with the cohomology group.} de Rham cohomology $H^{p}_{dR, M}$ on manifold $M$. An important cohomology that we have not explicitly introduced so far is \textbf{\v{C}ech cohomology}. We have, however, already used combinatorics of open covers $\U$ which is an important ingredient in \v{C}ech cohomology and sheaf cohomology \cite{gallier}. This cohomology provides details of the local features of topology, as we will see. In particular, we find that constant $\constant$ is a 2-cocycle in the de Rham-\cech double cohomology that stitches the total local and global data on manifold $M$. 
\begin{theorem}[Isomorphism between de Rham and \cech cohomology]\label{theorem}
    There exists an isomorphism between the classes of de Rham cohomology and \cech cohomology in the topological quantization.
\end{theorem}
This would imply that there is an isomorphism between classes of de Rham cohomology and \cech cohomology. The restriction map from de Rham classes to de Rham-\cech double cohomology classes and the restriction map from Cech classes to de Rham-\cech double cohomology classes are also interesting but will not covered in this paper.

\subsection{The Role of de Rham-\cech Cohomology}
In this subsection, we will see the roles played by the classes of de Rham-\cech cohomology in Dirac's quantization and local 1-form potentials $A_{\alpha}$. To do that, we must first motivate the definition of \cech cohomology in the present context.

A presheaf $\F$ is defined to be a contravariant functor from a category of open set on topological space $X$ to a category of abelian group
\begin{equation}
\F \colon \text{Open}(X) \rightarrow \text{Cat}(G)
\end{equation}
where Cat$(G)$ is an abelian category of $G$. Equivalently, we can say that a presheaf $\F$ associates an abelian group $G$ to $\U$ on $X$. We will make it more precise later. For $\U \subset \mathcal{V}$, we define a restriction map
\begin{equation}
 \rho^\U_\mathcal{V} \colon \F(\mathcal{V}) \rightarrow \F(\U)
 \end{equation} 
 where, for let say $\U \subset \mathcal{V} \subset \mathcal{W}$, $\rho^\mathcal{W}_\U = \rho^\mathcal{V}_\U \cdot \rho^\mathcal{W}_\mathcal{V}$ is satisfied.
 Moreover, we see that open covering on $X$ is also important for this definition, see \cite{bott1982differential,grothendieck1957quelques}. In fact, one could ask for the definition of $\F(\Ua \cap \Ub)$. Or more appropriate to the current discussion, the meaning of $\F(\Ua \cap \Ub \cap \Ug)$? Answers to such questions can answer what is the global and local properties of $X$. In other words, can we use a local function to determine a global solution on $X$? This seems to be related to our discussion of singularity in a global gauge potential and the necessity of using local 1-forms with gauge transformation between them in the defined overlap.

It is useful to us when we do triangulation of our manifold $X$. In this way, it enables us to define simplicial complexes. Mainly, we wish to define $p$-cochains that are related by a map (or morphism) $\delta$. The essential idea is to first define the $p$-simplex for open cover $\mathfrak{U} = \{ \U_i \}$. We will denote a vertex $a$ in $\Ua$, we can denote two vertices $a$ and $b$ for a non-empty overlap $\Ua \cap \Ub$ and connect it by an edge which is a $1$-simplex. For a triple finite intersection $\Ua \cap \Ub \cap \Ug$, we have $2$-simplices as a triangle. Similarly, we can draw $p$-simplex for any non-empty overlap $\Ua \cap \Ub \cap \cdots \cap \U_\sigma$. We will denote such repeated collection in every non-empty overlap of simplex as the nerve of $\mathfrak{U}$ as $N(\mathfrak{U})$. For a discussion on simplicial complex see \cite{croom1978simplicial}.

Now that we have introduced $p$-simplices and presheaf $\F$, we are good to talk about \cech cohomology. We will define p-cochains $C^p(\U, \F)$ as a linear combination generated from $p$-simplices. It is a must to write $\U$ while defining p-cochains to show the continuing dependency of the cover. Also, these p-cochains are being defined for some presheaf $\F$. More concretely, we can write
\begin{equation}
C^p(\U, \F) = \prod_{\a < \b < \g< \dots < \sigma} \F(U_{\a \b \g\cdots \sigma})
\end{equation}
where we have used our previous notation $U_{\a \b \g\cdots \sigma} = \Ua \cap \Ub \cap \cdots \cap \U_\sigma$. We will denote a co-boundary map $\delta$
\begin{equation}\label{delta}
\delta \colon \chain \rightarrow C^{p+1}(\U, \F).
\end{equation}
One can easily verify that just like $d^2=0$ for the de Rham complex, it is $\delta^2 = 0$ for this complex as well. A p-cocycle $Z^p(\U,\F)$ is defined to be a p-cochain which is trivial under the map $\delta$
\begin{equation}
\delta \colon Z^p(\U,\F) \rightarrow 0
\end{equation}
which is stronger when one assumes the Poincare lemma. Similarly, one says that in \eqref{delta}, $C^{p+1}(\U, \F)$ is the boundary cochain of $\chain$. Visually, the complex of cochains looks like
\begin{equation}
\xymatrix{
C^0(\U,\F) \ar[r]^\delta & C^1(\U,\F)  \ar[r] ^\delta & C^2(\U,\F) \ar[r] ^\delta & \cdots \cdots \ar[r] ^\delta & C^p(\U,\F) 
}
\end{equation}
and $\delta$ induces a cohomology $\check{H}^p(X)$ which is the \cech cohomology, which is a set of p-coycles {\it modulo} p-coboundaries. The cohomology group can be defined as independent of $\U$. It will be necessary for us to also define a map
\begin{equation}
\delta^{-1} \colon C^{0}(\U,\F) \rightarrow C^{-1}(\U,\F)
\end{equation}
but this would require a little more context as we will see. It is also required to introduce {\it partition of unity} functions $\{p_\a\}$ associated to each $\{\U_a \}$ with the following properties
\begin{enumerate}
    \item $p_\a \geq 0$
    \item $\sum_\a p_a = 1$
    \item There exists a compact support for $p_\a$ in cover $U_\a$.
\end{enumerate}

Given a \cech cohomology, we can extract information about the topology for some coefficients in a presheaf $\F$. So \cech cohomology must answer any topological corrections of kind eq.~\eqref{lagrangian}. It is possible now to form a new double cohomology of de Rham cohomology and \cech cohomology. This would give a precise answer to our quest of understanding the topological corrections in eq.~\eqref{lagrangian}. We can simply, and roughly, endow a differential form to each p-cochain $C^p(\U, \F)$. For example, $A_{\alpha}$ are 0-cochain and 1-form. Similarly, the transition functions are 1-cochains (0-form) and the constant $c_{\a \b \g}$ are 2-cocycles (0-form) because they are closed since $\delta^2 g_{\a \b} = 0$. It would be nice to neatly write them as ${\mathbf {C^p(\U,\F,\Omega^q)}}$, which would represent an object of p-cochain and q-form. Note that $A_{\alpha}$, $g_{\a \b}$, and $c_{\a \b \g}$ are defined for $\{\U_\a \}$ and their overlaps as discussed above. As we will argue below that $\{ A_{\alpha}, g_{\a \b}, c_{\a \b \g}\}$ are the most fundamental topological information about $S^2$ under our consideration of charge quantization. This is often called `topological quantization'.

In the claim that $\{ A_{\alpha}, g_{\a \b}, c_{\a \b \g}\}$ describes a monopole in $S^2$ with the relevant de Rham-\cech classes, what is the pre-sheaf $\F$ for $C^p(\U,\F)$? In gauge theory with $U(1)$ action, this presheaf is $\F=\Z$. This is evident from our flux through curvature $F$ in eq.~\eqref{flux}. This condition is somewhat also important for quantization in gauge theory \cite{shao2023whats}. We will fix the presheaf from now on as $\Z$ and only write it when needed.

Let us now review the whole sec.~\ref{1.1} on Wu-Yang quantization as viewed in de Rham-\cech cohomology group $\double$. We always stick with our definition of a good cover\footnote{It should be noted that there exists a refinement of cover with direct limit $\U$ and one can always define $C^p(\U,\F)$ without any dependency on the cover \cite{bott1982differential}.} $\mathfrak{U}=\{ \U_\a \}$. An initial question is if de Rham-\cech cohomology gives insights into the obstruction in defining a global $A$ without singularity. A local 0-cochain, $q$-form $\lambda \in C^0(\U,\Omega^q)$ can be defined globally if in the overlap one can define
\begin{equation}\label{32}
\lambda_\a - \lambda_\b = 0,\quad \lambda_\a - \lambda_\b \in C^1(\U,\Omega^p)
\end{equation}
which is to say that $\lambda_\a$ and $\lambda_\b$ are equivalent and extendible to each other's region. In this sense, the transition function $g_{\a \b} \colon \lambda_\a \rightarrow \lambda_\b$ becomes an identity map. While eq.~\eqref{32} is satisfied for 0-cochain and the closed form $F_\a \in C^0(\U,\Omega^2)$ and thus we can ignore the subscript\footnote{So that $F_\alpha = F_\beta$, which is not necessarily true for $A$.}, it is not satisfied for $A_\a$ because of a different cocycle condition. This means that \cech theory is important to understand the obstructions of defining a local theory globally.

Translating everything and finding the correction in $\int_{\Gamma} A$ is the goal here. (We will find the isomorphism between the de Rham classes and \cech classes in this exercise.) The gauge transformations $g_{\a \b}$ in sec.~\ref{1.1} are 0-form and 1-cochain ${\mathbf {C^1(\U,\F,\Omega^0)}}$. Now, a very simple claim is that the following are equivalent
\begin{enumerate}
    \item The operation $d\colon g_{\a \b} \rightarrow dg_{\a \b}$ in ${\mathbf {C^1(\U,\F,\Omega^1)}}$ where $dg_{\a \b} = A_\a - A_\b$.
    \item The operation $\delta \colon A_\a \rightarrow A_\a - A_\b$ which also lies in ${\mathbf {C^1(\U,\F,\Omega^1)}}$.
\end{enumerate}

What about $d\colon dg_{\a \b}$ and $\delta \colon g_{\a \b}$? The former vanishes because of $d^2=0$ while the latter are the objects in ${\mathbf {C^2(\U,\F,\Omega^0)}}$. We will denote those by $c_{\a \b \g}$ and is given by
\begin{equation}
\delta \colon g_{\a \b} \rightarrow \eta_{\a \b \g} = g_{\a \b} + g_{\b \g} + g_{\g \a}
\end{equation}
which is also our cocycle. Now, for convenience, we can create a table for all the classes.
\begin{table}[h]
    \label{tableone}
    \centering
    \begin{tabular}{ccccc}
        $\Omega^3$& 0&     \\
        $\Omega^2$& $F$&  0  & \\
         $\Omega^1$& $A_{\a}$& $dg_{\ab}$  &0 \\
         $\Omega^0$&  & $g_{\a \b}$ & $\eta_{\a \b \g}$ & 0 \\

       & ${\mathbf {C^0(\U,\F,\Omega^q)}}$ & ${\mathbf {C^1(\U,\F,\Omega^q)}}$ & ${\mathbf {C^2(\U,\F,\Omega^q)}}$ & ${\mathbf {C^3(\U,\F,\Omega^q)}}$\\
    \end{tabular}
    
    \caption{A table box for all the classes in our de Rham-\cech cohomology.}
\end{table}

\begin{center}
\begin{figure}
\begin{tikzcd}
    {H_{dR}} &&& {H^*(\mathbf{C}, \mathcal{U})} \\
    \\
    {H_{\text{\v{C}ech}}} &&& {H^*(\mathbf{C}, \mathcal{U})} \\
    {} &&& {} && {}
    \arrow["r", from=1-1, to=1-4]
    \arrow["g"', from=1-1, to=3-1]
    \arrow["{\mathbb{I}}"', from=3-4, to=1-4]
    \arrow["f"', from=3-1, to=3-4]
\end{tikzcd}
\caption{Here the maps $r$ and $f$ are restriction maps which maps de Rham cohomology and \cech cohomology to the double complex ${H^*(\mathbf{C}, \mathcal{U})}$.}
\end{figure}

\end{center}

In order to find the quantization condition, we again impose the consistency condition for any Euclidean propagator and get the similar cocycle condition Eq.~\eqref{consistency} and thus getting the quantization condition $\eta_{\a \b \g} = 2 \pi \epsilon _{\a \b \g}$. (One may similarly do other cases with QFTs, for example, WZW models \cite{alvarez1985topological} where one encounters closed 3-forms rather than 2-forms.)

Finally, we see that for locally defined 1-forms $A$, we get a cocycle condition that implies the quantization condition. From table.~\ref{tableone}, we can now find the isomorpism between de Rham classes and \cech classes. If we start with a globally defined 2-form $F$ in ${\mathbf {C^0(\U,\F,\Omega^2)}}$, we can find the locally defined constant $\eta$ in ${\mathbf {C^2(\U,\F,\Omega^0)}}$ using $d$ and $\delta$ (and their inverses) maps. In this way, we have the isomorphism between the de Rham cohomology and \cech cohomology. 


\subsection{Final Comment}
The exposition in the previous section suggests that there are quite many deep algebraic geometric structures in the example of Wu-Yang monopoles. There has been use of Deligne cohomology as well for studies like WZW \cite{Gawedzki:2002se}. 

We also see that presheafs and sheafication can give us an idea about how information works on a manifold. Indeed, one can use such algebraic geometry tools to study how local and global information appears for a certain gauge theory. A good example would be to understand Higgs bundle in this context. Moreover, Theorem \ref{theorem} is important from a mathematical perspective as well. In this document, we only emphasized the physical aspects from gauge theory side.

We note that most of this paper was written in 2023.

{\bf Aknowledgements$\colon$} I was fortunate to learn mathematics relevant to this paper from a lot of people and I thank them all for those discussions. In particular, I would like to thank A.K Maloo for teaching me a lot about abstract algebra.
\bibliography{local}
\bibliographystyle{utphys}

E-mail: \textsf{ aayushverma6380@gmail.com}
\end{document}